\documentclass[amsmath,nobibnotes,aps,superscriptadress,amssymb,prl,aps,showpacs,superscriptaddress,twocolumn]{revtex4-1}
\usepackage{amsmath,amsfonts,amssymb,amsthm,graphics,graphicx,epsfig,bbm}
\usepackage[colorlinks=true,citecolor=blue,linkcolor=blue,urlcolor=blue]{hyperref}
\usepackage[usenames]{color}
\usepackage{graphicx}
\usepackage{subfigure}
\usepackage{amsmath}
\usepackage{epsfig}
\usepackage{dcolumn}
\usepackage{bm}
\usepackage{color}
\usepackage{times}
\usepackage{epstopdf}
\usepackage{amssymb}
\usepackage{amstext}
\usepackage{latexsym}
\usepackage{hyperref}
\usepackage{amsfonts}
\usepackage{psfrag}
\usepackage{soul,xcolor}
\usepackage[normalem]{ulem}
\usepackage{dsfont}
\usepackage{txfonts}
\usepackage{physics}
\usepackage{footnote}

\begin{document}
\setstcolor{red}
	
\title{Towards Pricing Financial Derivatives with an IBM Quantum Computer} 
\date{\today}

\author{Ana Martin}
\thanks{Both authors have contributed similarly to this work.}
\affiliation{Department of Physical Chemistry, University of the Basque Country UPV/EHU, Apartado 644, 48080 Bilbao, Spain}
\affiliation{Department of Physics, Shanghai University, 200444 Shanghai, China}
	 
\author{Bruno Candelas}
\thanks{Both authors have contributed similarly to this work.}
\affiliation{Department of Physical Chemistry, University of the Basque Country UPV/EHU, Apartado 644, 48080 Bilbao, Spain}
	
\author{\'Angel Rodr\'iguez-Rozas}
\affiliation{Santander Analytics, Risk Division, Banco Santander, Avenida de Cantabria S/N, 28660 Boadilla del Monte, Madrid, Spain}

\author{Jos\'e D. Mart\'in-Guerrero}
\affiliation{IDAL, Electronic Engineering Department, University of Valencia, Avgda. Universitat s/n, 46100 Burjassot, Valencia, Spain}

\author{Xi Chen}
\affiliation{Department of Physics, Shanghai University, 200444 Shanghai, China}
\affiliation{Department of Physical Chemistry, University of the Basque Country UPV/EHU, Apartado 644, 48080 Bilbao, Spain}
	
\author{Lucas Lamata}
\affiliation{Department of Physical Chemistry, University of the Basque Country UPV/EHU, Apartado 644, 48080 Bilbao, Spain}

\author{Rom\'an Or\'us}
\affiliation{Donostia International Physics Center, Paseo Manuel de Lardizabal 4, 20018 San Sebasti\'an, Spain }
\affiliation{IKERBASQUE, Basque Foundation for Science, Maria Diaz de Haro 3, 48013 Bilbao, Spain}
\affiliation{Multiverse Computing, Pio Baroja 37, 20008 San Sebasti\'an, Spain}
	
\author{Enrique Solano}
\affiliation{Department of Physical Chemistry, University of the Basque Country UPV/EHU, Apartado 644, 48080 Bilbao, Spain}
\affiliation{IKERBASQUE, Basque Foundation for Science, Maria Diaz de Haro 3, 48013 Bilbao, Spain}
\affiliation{Department of Physics, Shanghai University, 200444 Shanghai, China}

\author{Mikel Sanz}
\email[Corresponding author: ]{\qquad mikel.sanz@ehu.es}
\affiliation{Department of Physical Chemistry, University of the Basque Country UPV/EHU, Apartado 644, 48080 Bilbao, Spain}

\begin{abstract}
Pricing interest-rate financial derivatives is a major problem in finance, in which it is crucial to accurately reproduce the time-evolution of interest rates. Several stochastic dynamics have been proposed in the literature to model either the instantaneous interest rate or the instantaneous forward rate. A successful approach to model the latter is the celebrated Heath-Jarrow-Morton framework, in which its dynamics is entirely specified by volatility factors. On its multifactor version, this model considers several noisy components to capture at best the dynamics of several time-maturing forward rates. However, as no general analytical solution is available, there is a trade-off between the number of noisy factors considered and the computational time to perform a numerical simulation. Here, we employ the quantum principal component analysis to reduce the number of noisy factors required to accurately simulate the time evolution of several time-maturing forward rates. The principal components are experimentally estimated with the $5$-qubit IBMQX2 quantum computer for $2\times 2$ and $3\times 3$ cross-correlation matrices, which are based on historical data for two and three time-maturing forward rates. This manuscript is a first step towards the design of a general quantum algorithm to fully simulate on quantum computers the Heath-Jarrow-Morton model for pricing interest-rate financial derivatives. It shows indeed that practical applications of quantum computers in finance will be achievable in the near future.
\end{abstract}
	
\maketitle

\section{Introduction}
In Finance, derivatives are contracts whose value derives from the value of an underlying financial asset or a set of assets, like an index, bonds, currency rates, stocks, market indices or interest rates. Typical financial derivatives contracts include forwards, futures, swaps (currency swaps or interest rate swaps), caps, floors, swaptions, among many others. They are typically used either to manage (mitigate) risk exposure (hedging), or for pure speculation. In case of pricing interest-rate financial derivatives under the risk neutral assumption~\cite{ShreveSCFII, BrigoMercurioBook}, it is crucial to model accurately the time-evolution of interest rates. Several stochastic dynamics have been proposed in the literature to model either the instantaneous interest rate $r(t)$ (also known as the instantaneous spot rate or, simply, as short rate) or the instantaneous forward rate, which is the forward rate at a future, infinitesimal period $(T,T+\delta t)$ forecasted at a previous time $t$, denoted by $f(t,T)$ \cite{BrigoMercurioBook}. Simple dynamics based on one or two noisy (random) factors for modeling both the short rate and the forward rates, have been proposed \cite{BrigoMercurioBook,KLRSS09,KT10,L16}. For short rates, one- and two-factor models became popular, such as the the Vasicek model, the Hull\&White model, the CIR (Cox-Ingesroll-Ross) model and its CIR++ extension, as of one-factor models, the Gaussian-Vasicek model and the Hull-White Two-Factor model, as of two-factor models. Furthermore, their corresponding algorithms are straightforward to implement. However, these models suffer from the strong requirements which arise from the necessity to calibrate to market data and to capture, at the same time, correlation and covariance structures from the time evolution of different forward rates. A highly successful approach proposed to overcome these constraints is the celebrated Heath-Jarrow-Morton (HJM) framework \cite{HJM1990, HJM1991, HJM1992, Jarrow2002}, which directly models the time evolution of forward rates. Indeed, the HJM model is a general family of models from which most of the aforementioned models may be derived \cite{BrigoMercurioBook}. Here, the dynamics is entirely specified by its volatility factors. Although general, there is a trade-off between the number of noisy factors considered and the computational time when executing the algorithm. Therefore, the computational power limits the accuracy of the model.

 Quantum computing (QC) has emerged in the last years as one of the most exciting applications of quantum technologies~\cite{Nielsen2000}, which promises to revolutionize the computational power at our disposal. In QC, entanglement, probably the most characteristic signature of quantum physics, is employed as an extra resource to speed up the performance of the computation, since it allows us to parallelize the calculations. Multiple algorithms with provable quantum speed up with respect to their best classical counterparts have been proposed for prime factorization \cite{Shor97}, searching in a list \cite{Grover01}, solving systems of linear equations \cite{HHL} or finding the largest eigenvalues and eigenvectors of a given matrix \cite{QPCA}, among many others. A particularly relevant application are quantum simulations, in which a controllable quantum system simulates the dynamics of another quantum system of interest whose classical simulation would be highly inefficient. Examples of applications of quantum simulations can be found in spin systems \cite{Laetal11,LHMLFWS14}, quantum chemistry \cite{G-ALHMSSL16,KMTTBCG17,A-LG-TSZPC18,BBMN18}, quantum field theories \cite{KDMMPSSLS18,P18}, fluid dynamics \cite{MSLESE15}, or quantum artificial life \cite{A-RSLS14,A-RSLS16,A-RSLS18}. Some applications of quantum technologies to finances have already been proposed \cite{RomanReview,Roman18,Rebentrost18,QuantumFinance}, but only few experiments have been carried out so far \cite{VK18,WE19}. State-of-the-art technology, however, only provides us with small noisy quantum chips, which limits the applicability of digital quantum simulations to toy models.

In this article, we employ an efficient quantum principal component analysis (qPCA) algorithm to effectively reduce the number of noisy factors needed to accurately simulate the joint dynamics of several time-maturing forward rates, according to the multi-factor HJM model. In addition, we implement this algorithm in the $5$-qubit IBMQX2 superconducting quantum processor of IBM. The volatility factors are estimated from $2\times 2$ and $3\times 3$ cross-correlation matrices between different time-maturing forward rates based on historical data. This is, to our knowledge, both the first quantum computing experiment in financial option pricing and the largest implementation of the qPCA algorithm on a quantum platform. Although for small matrices the problem can be easily solved on a classical computer, this contribution represents a first promising attempt towards the quantum computation of large-scale financial problems which are today prohibitively expensive. In the present Noisy Intermediate-Scale Quantum (NISQ) technology era \cite{P18}, we extend the applications of quantum computers to the field of finance, paving the way for achieving useful quantum supremacy/advantage in the following years.

\subsection{Financial Derivatives} 

A T-maturity zero-coupon bond (also known as a pure discount bond) is a contract that ensures its investor to accrue one unit of currency at time $T$ (its maturity), whose price at a previous time $t$ is denoted by $P(t, T)$. From this definition, it is clear that at expiry of the contract we must have $P(T, T) = 1$. This time-dependent curve represents a fundamental element in the theory of risk-neutral derivative pricing \cite{BrigoMercurioBook} and will extensively be used throughout this article. $P(t, T)$ is also known as the curve of discount factors, since it is employed to calculate the present value of future cash-flows. The inverse of this amount is called the capitalization factor, providing the capitalization of a present quantity to a future time.

In finance, the instantaneous interest rate $r_t$ (also known as the instantaneous spot rate, or simply as short rate) is the rate of return of a risk-free investment at time $t$ (for example, a U.S. treasury bond) (see Ref. \cite{BrigoMercurioBook}). This is also the interest rate applied when borrowing money from the money market and it is given as an annual percentage. We denote by $B(t)$ the time-$t$ value $B(t)$ of the money market account, defined as

\begin{equation}
B(t) = \exp \left(- \int_0^t r(s) ds \right).
\label{def:MoneyMarketAcc}
\end{equation}

Of course, if today is the time $t$, the value of $r(t)$ can be observed in the money market and therefore, it constitutes a known value. However, for future times $T>t$, $r(T)$ is uncertain and modeled through a stochastic process. In the risk-neutral framework, when using the money market account $B(t)$ as num\'eraire, the link between the short-rate and the zero-coupon is indeed materialized by the risk-neutral pricing formula

\begin{equation}
P(t, T) = \mathbb{E}^{\mathbb{Q}_B} \left[ \frac{B(t)}{B(T)} \times 1 \middle\vert \mathcal{F}_t \right] = \mathbb{E}^{\mathbb{Q}_B} \left[  e^{\left(- \int_t^T r(s) ds \right)} \middle\vert \mathcal{F}_t \right],
\label{eq:BondAsRate}
\end{equation}
where $\mathbb{Q}_B$ is the equivalent martingale measure associated to the numéraire $B(t)$, and $\mathcal{F}_t$ denotes the filtration of the information observed in the market until time $t$. Therefore, if today is the time $t$, then $P(t,T)$ is deterministic and it should match the information observed in the market. However, at any future time $t_f$ from today, $t < t_f < T$, $P(t_f, T)$ represents a random variable whose  value is model dependent.

Models for short-rate are typically classified depending upon the number of noisy factors that defines its dynamics. Popular one-factor short-rate models include the Vasicek, the Hull\&White model, the CIR (Cox-Ingesroll-Ross) model and its CIR++ extension, among others. They fast became of lesser interest due to their limitation when pricing financial instruments whose pay-offs involve the joint distribution of several of such rates at different maturities, mainly due to its incapability to exhibit the intrinsic decorrelation among them. Motivated from this observation, multi-factor models appeared to enrich the correlation structure. As a result, several two factor models were proposed, such as the Gaussian-Vasicek model and the Hull-White Two-Factor model.

Despite the freedom when modeling the instantaneous short-rate in the models mentioned above, some limitations may appear when attempting to calibrate a particular model to the current (observed) market curve of discount factors and to capture, at the same time, the correlation and covariance structure of forward rates. The first sound alternative to short-rate models was introduced by Heath, Jarrow and Morton in 1992 \citep{HJM1992}, developing a general framework for modeling the instantaneous forward rates. In its multi-factor version, determining the number of noisy factors needed becomes a trade-off between the ability, with increasing noisy factors, to better reproduce correlation and covariance structures while capturing market data, and the computational cost when performing a numerical simulation.

The connection between the forward rates $f(t, T)$ and the short rate $r(t)$ is established through the bond price as

\begin{equation}
f(t,T) = -\frac{\partial}{\partial T} \log P(t,T).
\label{eq:FwdAsBond}
\end{equation}

When $f(t, T)$ is known for all $T$, we must have

\begin{equation}
P(t,T) = e^{-\int_t^T f(t,s) ds}.
\label{eq:BondAsFwd}
\end{equation}

By differentiating Eq. (\ref{eq:BondAsRate}) with respect to $T$ we obtain

\begin{equation}
-\frac{\partial P(t,T)}{\partial T}
  = \mathbb{E}^{\mathbb{Q}_B} \left[ \exp \left(- \int_t^T r(s) ds \right) r(T) \middle\vert \mathcal{F}_t \right].
\label{eq:FwdAsExpectedRateQB}
\end{equation}

By changing to the $T-$forward measure $\mathbb{Q}_T$ associated to the bond price num\'{e}raire $P(t,T)$ we have

\begin{eqnarray}
-\frac{\partial P(t,T)}{\partial T}
     & = \mathbb{E}^{\mathbb{Q}_T} \left[ \exp \left(- \int_t^T r(s) ds \right) r(T) \frac{P(t,T)}{\exp \left(- \int_t^T r(s) ds \right)} \middle\vert \mathcal{F}_t \right] \nonumber \\
     & = P(t,T) \mathbb{E}^{\mathbb{Q}_T} \left[ r(T) \middle\vert \mathcal{F}_t \right],
\label{eq:FwdAsExpectedRateQTcalculation}
\end{eqnarray}
and therefore,

\begin{equation}
f(t, T) =  \mathbb{E}^{\mathbb{Q}_T} \left[ r(T) \middle\vert \mathcal{F}_t \right].
\label{eq:FwdAsExpectedRateQT}
\end{equation}

As such, in the HJM multi-factor model, the evolution of a risk-neutral zero-coupon bond price satisfies the following equation

\begin{equation}
dP(t,T) = P(t,T) \left\lbrace r(t) dt + \sum_{i=1}^N \left( \int_t^T \sigma_i (t,s) ds \right) dW_i(t) \right \rbrace,
\label{eq:HJMbond}
\end{equation}
where $dW_i, i=1,...,N$, are the uncorrelated Brownian increments associated to the volatilities $\sigma_i$. Using the bond price dynamics (\ref{eq:HJMbond}) and (\ref{eq:FwdAsBond}), we have
\begin{equation}
d f(t,T) = \alpha(t,T)dt + \sum_{i=1}^N \sigma_i (t,T) dW_i(t),
\end{equation}
where

\begin{equation}
\alpha(t,T) = \sum_{i=1}^N \sigma_i (t,T) \int_t^T \sigma_i (t,s) ds.
\end{equation}

This unique choice for $\alpha(t,T)$ as a function of the volatility terms is what prevents arbitrage. As mentioned before, this is a general framework from which many short-rate models may be derived, upon the particular choices for the $\sigma$ terms. However, not every choice generates a Markovian dynamics. They must also be carefully selected in order to derive practical algorithms that are efficient in terms of computational times. One possibility that ensures Markovianity is to assume that the volatility factors only depend on the time to maturity, so $\sigma_i = \bar{\sigma_i} (T-t) = \bar{\sigma_i} (\tau)$. At this point, we can use time series data to calculate the functions $\bar{\sigma_i}$.  For this purpose, we build the covariance matrix between the changes in the forward rates for different time-maturities $\tau_j$ (typically for maturities at 1m, 3m, 6m, 1y, 2y, ...).  The result is a symmetric matrix whose diagonal terms are the variances of the rates, while the off-diagonal terms represent the covariances between each pair of rates.

Considering all possible time-maturing forwards is computationally costly for the numerical simulations. Using PCA we can obtain the most relevant eigenvectors and their associated eigenvalues. As seen in the literature, most of the evolution of the curve can be explained by considering 2 or 3 of such factors. Typically, it is observed that whenever the entries of the first principal component are all similar, then the dominant movement of the  curve will be a parallel shift. Also, the second component typically account for a twist in the curve. In general, if the eigenvalues are $\lambda_i$ and the eigenvectors are $\bold{v_i}$, the volatility factors will be given by

\begin{equation}
\bar{\sigma_i} (\tau_j) = \sqrt{\lambda_i} (\bold{v_i})_j.
\end{equation}

To illustrate our qPCA algorithm, we will apply this technique to the covariance matrix appearing in Fig.~$19.3$ in Ref.~\cite{Wilmott}, based on historical data for one-, three- and six-month rates. The matrix is

\begin{equation}\label{eq:sigma_3}
\sigma_3 = \begin{pmatrix}
0.000189 & 0.000097 & 0.000091 \\
0.000097 & 0.000106 & 0.000101 \\
0.000091 & 0.000101 & 0.000126 
\end{pmatrix}
\end{equation}

\subsection{Quantum Principal Component Analysis} 
Principal component analysis (PCA) is a mathematical technique which allows us to find the optimal low-rank approximation of a given matrix by computing its spectral decomposition in eigenvalues and eigenvectors. Indeed, this approximation discards the smallest eigenvalues of the matrix, keeping only the principal components of the spectral decomposition. This technique is of paramount importance for a variety of applications but, unfortunately, the computational cost is too high when the size of the matrix is large. It is in this context in which quantum algorithms and quantum computers may play a relevant role. Indeed, in Ref. \cite{QPCA} the authors provided an elegant quantum algorithm to perform PCA with an exponential speed-up. The authors assume that the matrix can be represented by a quantum state, i.e. it is a non-negative matrix with trace equal to one, which covers a wide range of interesting cases, including the case under study in this manuscript of covariance matrices associated to volatilities. 
In this manuscript, we implement a slightly modified version of the aforementioned algorithm, which is better adapted to be run in a small and noisy quantum chip, typical in this NISQ technology era \cite{Pr18}.This allows us to reduce the number of noisy presented within the HJM model. This is the first step toward the construction of general Quantum Computing algorithm to fully simulate the HJM model on the IBM Quantum Computer for pricing interest rate financial derivatives. In the following Section, we briefly describe the algorithm. 
\begin{figure*}[t]
\centering
\includegraphics[width=\textwidth]{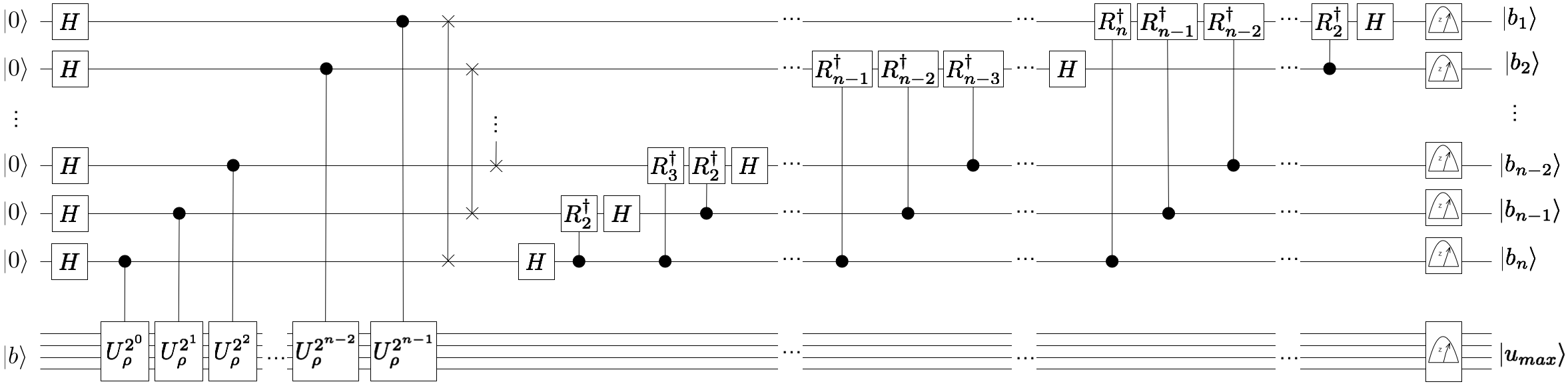}
\caption{{\bf Quantum circuit implementation for $n+\log N$ qubits.} \small The first $n$ qubits are dedicated to the binary codification of the maximum eigenvalue of the matrix $\sigma_N$ and they are initialized in the site $|0\rangle$. The rest of the qubits, a total of $\log N$, encodes the estimation of the corresponding eigenvector and are initialized on a random state $|b\rangle$. The single qubit gate $H$ correspond to the Hadamard gate. The rest of the gates are controlled operations. The controlled $U_\rho^{2^k}$ gate applies the matrix $U=e^{i t \sigma_N}$ $2^k$ times on the last set of qubits. The controlled $R^\dagger_k$ gate applies the matrix 
$\begin{tiny}
\begin{pmatrix}
1&0\\
0&e^{2\pi i/2^k}
\end{pmatrix}
\end{tiny}$ 
to each target qubit. After performing all the operations, if the initial $|b\rangle$ is appropriate,  one get the final state $|b_1b_2...b_n\rangle\otimes |u_{max}\rangle$. Where $|b_1b_2...b_n\rangle$ is the $n$-bit estimation of the eigenvalue $\lambda_j^{(n)}$ and $|u_{max}\rangle$ is the best estimation of the exact eigenvector of $\sigma_N$.}\label{fig:QPCA_N}
\end{figure*}
\section{Quantum Circuit}
Let us consider a non-negative matrix $\sigma_N \in \mathbbm{R}^N \times \mathbbm{R}^N$ with $\text{tr}[\sigma_N] = 1$, which is the matrix whose principal components we want to compute. Let us assume that we can efficiently generate the unitary $e^{i t \sigma_N}$. It has been proven in the literature that, under certain conditions such as sparsity of the matrix or the access to several copies of $\sigma_N$, this is possible. This matrix admits a spectral decomposition $\sigma_N = \sum_{j=1}^N \lambda_j | u_j\rangle\langle u_j|$, with $0\leq \lambda_j\leq 1$ and $\sum_{j=1}^N \lambda_j = 1$, and we assume that $\sigma_N$ can be very well approximated by a matrix $\rho_r = \sum_{j=1}^r \lambda_j |u_j\rangle \langle u_j |$ with rank $r\ll N$. Therefore, the goal of the algorithm is the determination of the $r$ largest eigenvalues of $\sigma_N$ and their corresponding eigenvectors. If we want to determine the eigenvalues with an $n$-bit precision, we will need $n+ \log N$ qubits, as depicted in Fig.~\ref{fig:QPCA_N}, which represents the gate decomposition of the algorithm. A priori, we do not know the eigenvectors of our algorithm. Hence, we cannot make use of quantum phase estimation to compute directly the corresponding eigenvalue. Consequently, we initialize our system in a random state $|b\rangle$ whose (unknown) decomposition in terms of the eigenbasis is given by $|b\rangle = \sum_{j=1}^N \beta_j |u_j\rangle$. If we take a random vector, the probability that there exists a component $\beta_k = 0$ is zero. The quantum state after the quantum Fourier transform can be written as $|\Psi_b\rangle = \sum_{j=1}^N \beta_j |\lambda_j^{(n)} \rangle \otimes |u_j\rangle$, so eigenvalues and eigenvectors are entangled. However, if our assumption that $\sigma_N$ is well approximated by the $r$-rank matrix $\rho_r$ is correct, then the highest eigenvalues should be around $1/r\approx \sum_{k=1}^n y_k 2^{-k}$. Calling $|y^{(n)}\rangle = |y_1\, y_2 \ldots y_n\rangle$ the vector of these components, it means that, by projecting the eigenvalue component $|\lambda_{j}^{(n)}\rangle$ of the state $|\Psi_b\rangle$ around this component, one may obtain the eigenvector corresponding to the maximum eigenvalue, i.e. $\langle y^{(n)}| \otimes \mathbbm{1} |\Psi_b\rangle \approx |u_{\text{max}}\rangle$. It is possible, especially when $n$ is small, as may happen in the NISQ chips, that the $n$-bit approximation of the eigenvalue cannot be able to distinguish between two or more eigenvectors. In this case, the projection is not into the maximum eigenvalue, but into a $K$-dimensional subspace containing the indistinguishable components $\langle y^{(n)}| \otimes \mathbbm{1} |\Psi_b\rangle = \sum_{j=1}^K \tilde{\beta}_j |u_{j}\rangle$, where the $\tilde{\beta}_j$ are the normalized $\beta_j$ in the subspace. As we do not know a priori whether $K>1$ or not, we could start with a different random state $|c\rangle = \sum_{j=1}^N \gamma_j |u_j\rangle$, which leads to $|\Psi_c\rangle = \sum_{j=1}^N \gamma_j |\lambda_j^{(n)} \rangle \otimes |u_j\rangle$. After projecting into $|y^{(n)}\rangle$ the expect state is a different superposition $\sum_{j=1}^L \tilde{\gamma}_j |u_{j}\rangle$ with high probability, which helps us to check whether we have actually identified the eigenvector corresponding to the maximum eigenvalue. Otherwise, we must increase the $n$-bit precision until a unique eigenvalue is identified.

Let us assume now that the $n$-bit precision is sufficient to determine a unique eigenvector. Taking into account the constraints due to the small number of qubits and the noise of the chip and the operations, we can sequentially improve the result of the eigenvector. As described above, we start the protocol with a random quantum state $|b_0\rangle$, to which the noisy algorithm is applied and the projection into the $|y^{(n)}\rangle$ subspace is performed. Let us call the result $|\Psi_{b_0}\rangle$, which is an approximation for the eigenvector. If we employ now this state as initial state in the protocol, $|\Psi_{b_0}\rangle = |b_1\rangle$, then one expects that the approximation for the eigenvector provided by $|\Psi_{b_1}\rangle$ improves the fidelity due to the cancelation of coherent errors associated to the $\beta$ components. Nonetheless, there is a limitation in this sequential improvement related to the decoherence of the qubits and the statistical error of the measurement. In any case, the result can be (slightly) further improved by performing measurements in different bases and averaging, since this cancels some systematic errors of the gates. 

\section{Results}
\begin{figure}[t]
\centering
\includegraphics[width=0.48\textwidth]{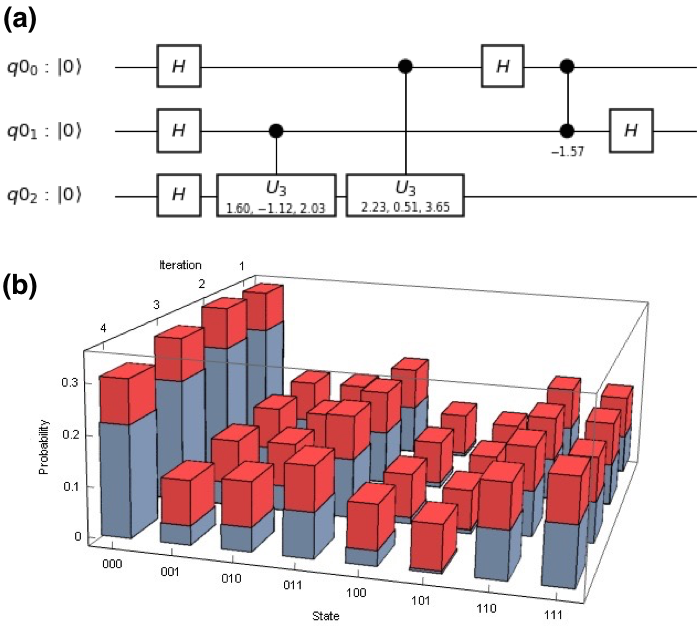}
\caption{{\bf (a) Quantum circuit implementation for the $2\times 2$ matrix.} {\small The first two qubits encodes the $2$-bit estimation of the greatest eigenvalue of $\rho_2$ and are initialized on the state $|0\rangle$. The last qubit is dedicated to the estimation of the corresponding eigenvector. It is initialized on a random state $|b\rangle$. For the first iteration, we have initialized it on the state $|+\rangle$. The single qubit gates represented by the letter $H$ refers to the Hadamard gate. The controlled $U_3$ gates represent the unitary controlled operations called $U_\rho^{2^k}$ on Fig \ref{fig:QPCA_N}. The last two qubit gate is a controlled $S^\dagger$ gate, controlling the first qubit and acting on the second. It is represented by a dot on each qubit. The final state of the system after running the circuit and taking measures is $|11\rangle\otimes(0.719|0\rangle+0.659|1\rangle)$}. {\bf (b) Populations for each iteration.} {\small The graphic shows the experimental probabilities of finding the three qubits in each state and its corresponding errors (red bars) for the four iterations of the algorithm. We have considered both statistical and experimental errors, assuming for the latter an error of $8\%$ for each two-qubit gate.}}
\label{fig:example1_iteration1}
\label{graph1}
\end{figure}

As described in the previous section, the protocol is divided into two parts. Firstly, we estimate the eigenvector $|u_\text{max}\rangle$ corresponding to the largest eigenvalue $\lambda_\text{max}$. We start with a random state, apply the circuit implementation shown in Fig.~\ref{fig:example1_iteration1}, project on the binary $n$-bit estimation for the largest eigenvalue $|y^{(n)}\rangle$, and use this state as initial state of the process, which sequentially approaches the exact eigenvector. Afterwards, we use this eigenvector to get a more accurate approximation for the eigenvalue $\lambda_\text{max}$ by means of quantum phase estimation. 

For the estimation of the eigenvector, we start with a random state $|b_0\rangle$. Hence, the initial state of the system is $|0\rangle\otimes|0\rangle\otimes|b_0\rangle$. After the first iteration and projecting on the computational basis the eigenvector, we obtain a first estimation, which we will call $|b_1\rangle$, and use it as the initial state of the system on the next iteration. This is: $|0\rangle\otimes|0\rangle\otimes|b_1\rangle$. We will continue this process and iterate $k$ times until $|b_{k-1}\rangle\approx|b_k\rangle$. Once we reach that point, we can say that $|b_k\rangle\approx|u\rangle$.

Let us now estimate the eigenvalue $\lambda_\text{max}$. Once the first part is finished and we have an accurate approximation for $|u_\text{max}\rangle$, we can apply quantum phase estimation \cite{Nielsen2000} to obtain $\lambda_\text{max}$ with $n$-bit precision. The precision is limited in this case by the size of the processor. Our aim is to apply the algorithm to the $3\times 3$ matrix given in Eq. \eqref{eq:sigma_3} in the $5$-qubit IBMQX2 quantum processor. Firstly, we will solve the $2\times 2$ submatrix of $\sigma_3$ containing only two maturities, and afterwards, we will solve the $4\times 4$ expansion of the same matrix. Despite the small size of the problem, the volume of the quantum algorithms allowed in this processor is almost achieved, but we can still obtain relatively accurate results. We have run the algorithm in both the simulator provided by QISKIT and the real IBM quantum processor, reaching accurate results in both cases. 
 
\subsection{$2\times 2$ matrix}

Firstly, we need to codify the covariant matrix in a quantum state, so we only need to normalize it with respect to its trace,
\begin{equation}
\rho_2=\frac{\sigma_2}{\tr(\sigma_2)} = \begin{pmatrix}
0.6407&0.3288\\
0.3288&0.3593
\end{pmatrix},
\end{equation}
whose spectral decomposition is given by
\begin{eqnarray}
\lambda_1=0.8576 \qquad&|u_1\rangle=0.8347|0\rangle+0.5508|1\rangle\label{eq:rho2eigenvector1},\\
\lambda_2=0.1424 \qquad&|u_2\rangle=0.5508|0\rangle-0.8347|1\rangle.
\end{eqnarray}
Let us remark that $\lambda_\text{max} \gg \lambda_2$, a usual characteristic of these correlation matrices, so we can apply the PCA technique to find the optimal low-rank approximation of $\rho_2$. Let us now define the unitary
\begin{equation}
U_{\rho_2}=e^{2 \pi i \rho_2}=\begin{pmatrix}
0.6260- 0.3068 i &-0.7170 i\\
-0.7170 i & 0.6260+ 0.3068 i 
\end{pmatrix}.
\end{equation}

For the first part of the protocol, we will make use of three qubits, two for a $2$-bit approximation of the eigenvalue, and a third one one to represent the eigenvector. We apply the first part of the protocol as described above, starting with a quantum state $|b_0\rangle = \frac{1}{\sqrt{2}}(|0\rangle + |1\rangle)$ and projecting into the $|y^{(n)}\rangle = |11\rangle$ state. After the $4$th iteration, each of them averaged over 8192 realizations, the outcome vector estimating the eigenvector stabilizes and we stop. With this final eigenvector, we also rotate the measurement basis in $x$, $y$ and an arbitrary direction $r = (\cos \alpha , -e^{i\beta}\sin \alpha; e^{i\beta}\sin \alpha, e^{i\gamma}\cos \alpha)$, with $\alpha = 1.00$, $\beta = 0.80$ and $\gamma = 0.16$, to compute the relative phase and to improve the accuracy of the solution provided. Our estimation for $|u_\text{max}\rangle$ is consequently given by
\begin{eqnarray} \label{maxeigvec}
|u_\text{max}\rangle &=& \left [ (0.87 \pm \delta) -i(0.10\pm \delta)\right ] |0\rangle \nonumber \\
&+& \left((0.47 \pm \delta)+i(0.10\pm \delta)\right) |1\rangle,
\end{eqnarray}
with $\delta = 0.9$ the error estimated from the qubit and measurement fidelity provided by IBM and the statistical error related to the number of repetitions. Let us remark that we have split the complex phase between both states using the global phase. The estimation for the coefficients after each iteration in the $z$ basis is provided in Table \ref{tab:coef2}. We can observe that the algorithm has already converged in the first iteration, and the variations are within the estimated error. We take the eigenvector produced in the last iteration and repeat the algorithm with this one as initial state measuring in $x$, $y$ and a $r$-random direction to check possible relative phases and to try to remove systematic errors, which yields the states $|b_x\rangle = 0.878 |0 \rangle +(0.421+i 0.230)|1\rangle$, $|b_y\rangle = 0.878|0\rangle +(0.427+i 0.220)|1\rangle$, and $|b_r\rangle = 0.985|0\rangle +0.175|1\rangle$.
\begin{table}[h!]
  \begin{center}
    \caption{Estimated coefficients of the eigenvector for consecutive iterations of the algorithm in modulus and measured in the $z$ basis. Here, the state of the previous iteration is employed as initial state in the following iteration until the values are stabilized. Measurements of the eigenvector are performed in the $z$ basis and repeated for 8192 realizations.}
    \label{tab:coef2}
    \begin{tabular}{c|c|c}
      \textbf{Iteration} & $c_0^z$ & \hspace{0.4cm}$c_1^z$ \\
      \hline
      1 & \hspace{0.4cm}0.719\hspace{0.4cm} & \hspace{0.4cm}0.695\hspace{0.4cm}\\
      2 & \hspace{0.4cm}0.707\hspace{0.4cm} & \hspace{0.4cm}0.707\hspace{0.4cm}\\
      3 & \hspace{0.4cm}0.720\hspace{0.4cm} & \hspace{0.4cm}0.694\hspace{0.4cm}\\
      4 & \hspace{0.4cm}0.680\hspace{0.4cm} & \hspace{0.4cm}0.734\hspace{0.4cm}\\
    \end{tabular}
  \end{center}
\end{table}

Let us remark that the previous estimation of the eigenvector was performed by projecting into the subspace estimating the eigenvalue into the two bit string $\lambda_\text{max}=0.11$. However, we can now apply quantum phase estimation to improve the estimation for the eigenvalue. We divide the problem into these two stages for two reasons. Firstly, we do not know a priori the value of the maximum eigenvalue only the approximate rank, and hence a low $n$-bit approximation covers a larger range, as explained in the previous section. Additionally, we observe a lower error when protocol performed in this manner, probably due to the accumulation of two qubit gates and the error in the projection for the eigenvalue estimation. However, we cannot be sure, since IBM does not provide the exact quantum circuit which they are performing in the processor. 

\begin{figure}
\centering
\includegraphics[width=0.48\textwidth]{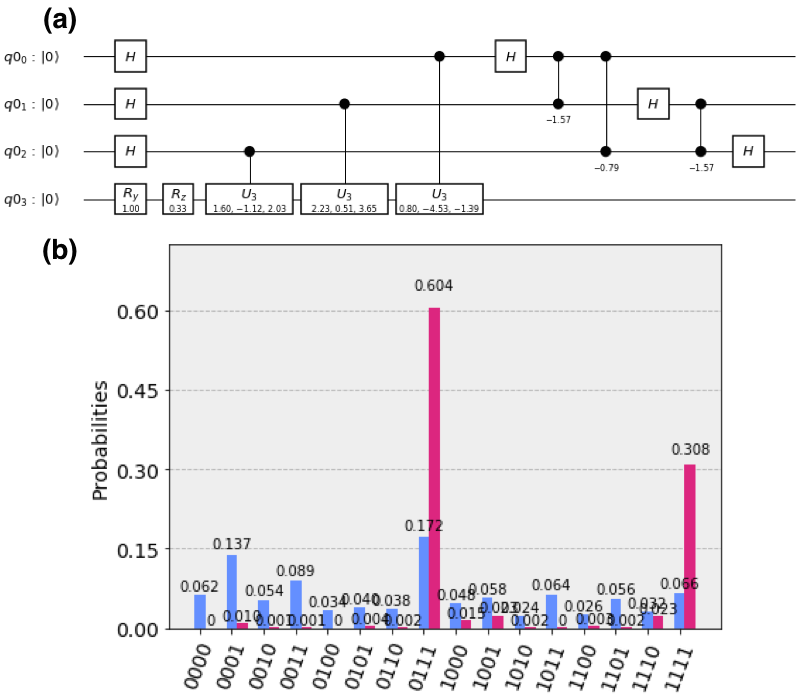}
\caption{{\bf (a) Quantum phase estimation circuit implementation for the $3$-bit estimation of the greatest eigenvalue, $\lambda^{(3)}$.} {\small After estimating the eigenvector $|u_{max}\rangle$, it is used to improve the estimation of the corresponding eigenvalue $\lambda$ by applying the quantum phase estimation algorithm. In this case we dedicate $3$ qubits to the binary codification of $\lambda$. The fourth qubit is initialized on the estimated eigenvector $|u_{max}\rangle$. The fifth two-qubit gate represents a controlled $T^\dagger$ gate, controlling the first qubit and acting on the third one. The rest of the gates are the same that have been applied on the previous part of the algorithm. Finally, one takes measures and gets the final state of the system: $|\lambda^{(3)}\rangle\otimes|u_{max}\rangle=|111\rangle\otimes(0.808|0\rangle+0.600|1\rangle)$}. {\bf (b) Populations of the $3$-bit eigenvalue estimation.} {\small This chart shows the probabilities of each state in simulator QISKIT (pink bars) and in the real quantum processor (blue bars) for the quantum phase estimation algorithm taking the previously obtained eigenvector given in Eq.~\eqref{maxeigvec}. The first qubit refers to the subspace of the eigenvector estimation. The next three qubits refers to the subspace of the binary estimation of the eigenvalue. The quantum circuit comprises at least $6$ entangling gates, which leads the system to an almost total decoherence, as reflected in the homogeneous distribution of probabilities in the real chip.}}
\label{fig:2x2QPhE}
\label{graph1}
\end{figure}

Let us now use three qubits for the eigenvalue estimation $\lambda_\text{max}=0.b_1b_2b_3$, keeping one qubit to encode the corresponding eigenvector. The depth of the circuit implementation grows and leads us to the decoherence of the system when we run it on the real quantum processor, as depicted in Fig. \ref{fig:2x2QPhE}. However, the result provided by the QISKIT simulator, produce the quantum state $|111\rangle\otimes\left[(0.808|0\rangle+0.600|1\rangle\right]$, which is an almost ideal result for the  $3$-bit string estimation of the eigenvalue. Indeed, the predicted eigenvalue is $\lambda=0.111$ in binary representation and corresponds to the number $\lambda = 0.875$ and the fidelity between $|u_\text{max}\rangle$ and the one obtained after performing the quantum phase estimation in the QISKIT simulator $|u_\text{QPE}\rangle$ is 
\begin{equation}
F=|\langle u_\text{QPE}| u_\text{max}\rangle|^2 = 0.977.
\end{equation}
This shows that, with few improvements in the gates and chips or with a lower level programming in the chip, one could substantially improve the results.

\subsection{$4 \times 4$ matrix}

\begin{figure}
\centering
\includegraphics[width=0.48\textwidth]{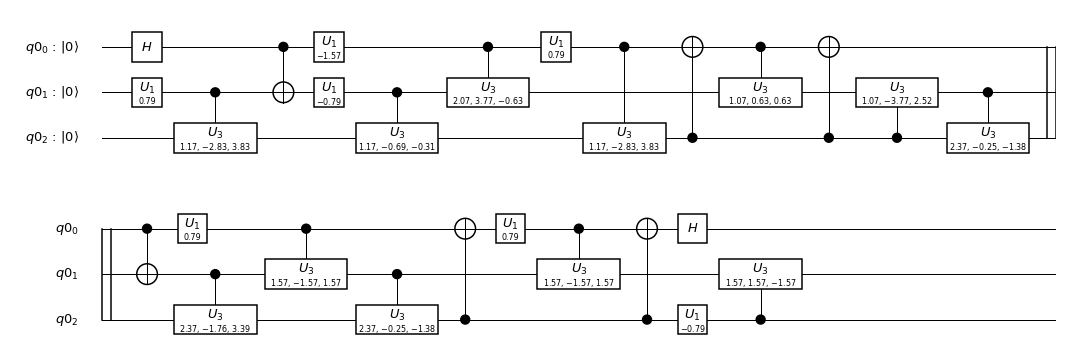} 
\caption{{\bf Quantum circuit implementation for the $4\times 4$ matrix.} {\small The first qubit is the only one dedicated for the binary codification of the greatest eigenvalue $\lambda$ of the matrix $\rho_4$. It is initialized on the state $|0\rangle$. The last two qubits encodes the estimation of the corresponding eigenvector and are initialized on a ransom state $|b\rangle$. The final state of the system after measuring is $|1\rangle\otimes|u_{max}\rangle$}}\label{fig:4x4FirstPart}
\end{figure}

In this case, the matrix $\sigma_3$ will be represented by the two-qubit quantum state
\begin{equation}
\rho_4=\frac{\sigma_4}{\tr(\sigma_4)}=\begin{pmatrix}
0.4489 & 0.2304 & 0.2162 & 0\\
0.2304 & 0.2518 & 0.2399 & 0\\
0.2162 & 0.2399 & 0.2993 & 0\\
0 & 0 & 0 & 0
\end{pmatrix}
\end{equation}
Thus, the unitary generated, $U_{\rho_4}=e^{2\pi i\rho_4}$, is given by
\begin{equation*}
\begin{small}
U_{\rho_4}=\begin{pmatrix}
0.415+0.048 i & -0.108-0.566 i & -0.029-0.702 i & 0\\
-0.108-0.566 i & 0.744-0.030 i & -0.285-0.181 i& 0\\
-0.029-0.702 i & -0.285-0.181 i & 0.618+0.099 i& 0\\
0 & 0 & 0 & 1
\end{pmatrix}.
\end{small}
\end{equation*}
The spectral decomposition of $\rho_4$, for the sake of comparability, is given by
\begin{eqnarray*}
\lambda_1=0.000 \qquad  |u_1\rangle &=& (0.000, 0.000, 0.000, 1.000),\\
\lambda_2=0.031 \qquad  |u_2\rangle &=& (-0.119, 0.786, -0.607, 0.000),\\
\lambda_3=0.169 \qquad  |u_3\rangle &=& (0.734, -0.342, -0.587, 0.000),\\
\lambda_4=0.800 \qquad  |u_4\rangle &=& (0.669, 0.516, 0.536, 0.000),
\end{eqnarray*}
where the vectors are expressed in the basis $\{|00\rangle, |01\rangle, |10\rangle, |11\rangle\}$. This problem is much more complicated than the previous one, since we do not implement $U_{\rho_4}$, but the controlled $U_{\rho_4}$. This matrix must be decomposed in terms of two-qubit gates, which dramatically increases the depth of the algorithm and, consequently, the decoherence and the errors. The quantum circuit implementation for this problem is shown in figure Fig \ref{fig:4x4FirstPart}. Following the aforementioned protocol, we start with the state $|b_0\rangle=\left(|00\rangle+|01\rangle+|10\rangle+|11\rangle\right)/2$ and provide the coefficients in the $z$ basis, for both the simulator and the real processor, in Table \ref{tab:coef3}.
\begin{table}[h!]
  \begin{center}
    \caption{Estimated coefficients of the eigenvector for consecutive iterations of the algorithm in modulus and measured in the $z$ basis. Here, the state of the previous iteration is employed as initial state in the following iteration until the values are stabilized. Measurements of the eigenvector are performed in the $z$ basis and repeated for 8192 realizations.}
    \label{tab:coef3}
    \begin{tabular}{c|c|c|c|c}
      \textbf{Iteration} & $c_{00}^z$ (chip) & $c_{01}^z$ (chip)& $c_{10}^z$ (chip)& $c_{11}^z$ (chip)\hspace{0.4cm}\\
      \hline
      1 & \hspace{0.4cm}0.542\hspace{0.4cm} & \hspace{0.4cm}0.503\hspace{0.4cm} & \hspace{0.4cm}0.466\hspace{0.4cm} & \hspace{0.4cm}0.487\hspace{0.4cm} \\
      2 & \hspace{0.4cm}0.531\hspace{0.4cm} & \hspace{0.4cm}0.498\hspace{0.4cm} & \hspace{0.4cm}0.493\hspace{0.4cm} & \hspace{0.4cm}0.477\hspace{0.4cm} \\
      3 & \hspace{0.4cm}0.543\hspace{0.4cm} & \hspace{0.4cm}0.493\hspace{0.4cm} & \hspace{0.4cm}0.494\hspace{0.4cm} & \hspace{0.4cm}0.468\hspace{0.4cm} \\
      4 & \hspace{0.4cm}0.502\hspace{0.4cm} & \hspace{0.4cm}0.492\hspace{0.4cm} & \hspace{0.4cm}0.523\hspace{0.4cm} & \hspace{0.4cm}0.482\hspace{0.4cm} \\
    \end{tabular}
      \end{center}
  \begin{center}
    \begin{tabular}{c|c|c|c|c}
      \textbf{Iteration} & $c_{00}^z$ (sim) & $c_{01}^z$ (sim) & $c_{10}^z$ (sim) & $c_{11}^z$ (sim)\\
      \hline
      1 & \hspace{0.4cm}0.719\hspace{0.4cm} & \hspace{0.4cm}0.695\hspace{0.4cm} & \hspace{0.4cm}0.695\hspace{0.4cm} & \hspace{0.4cm}0.695\hspace{0.4cm}\\
      2 & \hspace{0.4cm}0.707\hspace{0.4cm} & \hspace{0.4cm}0.707\hspace{0.4cm} & \hspace{0.4cm}0.695\hspace{0.4cm} & \hspace{0.4cm}0.695\hspace{0.4cm}\\
      3 & \hspace{0.4cm}0.720\hspace{0.4cm} & \hspace{0.4cm}0.694\hspace{0.4cm} & \hspace{0.4cm}0.695\hspace{0.4cm} & \hspace{0.4cm}0.695\hspace{0.4cm}\\
      4 & \hspace{0.4cm}0.680\hspace{0.4cm} & \hspace{0.4cm}0.734\hspace{0.4cm} & \hspace{0.4cm}0.695\hspace{0.4cm} & \hspace{0.4cm}0.695\hspace{0.4cm}\\
    \end{tabular}
      \end{center}
\end{table}
Afterwards, we measure in different bases in order to compute the relative phases, and take the average to cancel systematic errors. The estimation of the eigenvector is, therefore,
\begin{eqnarray}
|u_\text{max}\rangle&=&(0.6287+i 0.3991)|00\rangle+(0.4010+i 0.0693 i)|01\rangle+\nonumber\\
&&+(0.4807-i 0.1964)|10\rangle+(0.0305+i 0.0959)|11\rangle. \nonumber
\end{eqnarray}
The number of entangling gates performed for this algorithm is at least $18$, so the total estimated error $\delta$ in the coefficients, assuming the $8\%$ error per gate observed in the previous section, is over $100\%$, which makes in principle the result meaningless. 
\section{Conclusions}
We have proposed and implemented an efficient quantum algorithm to reduce the number of noisy factors present in the time evolution of forward rates according to the multi-factor Heath-Jarrow-Morton model. Indeed, this model considers several noisy components to accurately describe the dynamics of several time-maturing forward rates, which can be gathered in a cross-correlation matrix. The eigenvectors corresponding to the largest eigenvalues of this matrix provide the principal components of the correlations. When the considered data is large, this calculation turns out to be challenging. The principal components are experimentally estimated using a hybrid classical-quantum algorithm with the $5$-qubit IBMQX2 quantum computer for $2\times 2$ and $3\times 3$ cross-correlation matrices, which are based on historical data for two and three time-maturing forward rates. We have obtained a reasonable approximation for both the maximum eigenvalue and its corresponding eigenvector in the $2\times 2$ case. For the $4 \times 4$ matrix, the depth of the algorithm is too high and the experimental errors in the quantum processor prevent us from extracting any useful information. Simultaneously, the simulation in QISKIT shows that it would be achievable in a better experimental set. This means that we have exhausted the computational power provided by the current quantum processor in terms of gate fidelities, connectivity, and number of qubits. The main drawback is that cloud quantum computers force us to perform only black-box high-level programming. Therefore, if a lower level programming were available, the optimization of the quantum algorithm adapted to the chip constraints would allow us for dramatically increasing the algorithm volume. Nonetheless, this manuscript is a first step towards the design of a general quantum algorithm to fully simulate on quantum computers the HJM model for pricing interest-rate financial derivatives, and shows that practical applications of quantum computers in finance will be achievable even in the NISQ technology era.

The authors acknowledge the use of IBM QISKIT for this work. The  views  expressed  are  those  of  the  authors and do not reflect the official policy or position of IBM.  We also acknowledge support from Spanish Ram\'on y Cajal Grant RYC-2012-11391 and RYC-2017-22482, Basque Government IT986-16, the projects QMiCS (820505) and OpenSuperQ (820363) of the EU Flagship  on Quantum Technologies, Shanghai Municipal Science and Technology Commission (18010500400 and 18ZR1415500), and the Shanghai Program for Eastern Scholar. This material is also based upon work supported by the U.S. Department of Energy, Office of Science, Office of Advance Scientific Computing Research (ASCR), under field work  proposal number ERKJ335.

\end{document}